\documentclass[11pt]{article}
\usepackage{amsmath,amssymb,amsthm,amsxtra,bbm,bbold,bm,epsfig,multirow,longtable}
\usepackage{ulem,subfigure,hyperref,cite}
\usepackage[dvipsnames, table,xcdraw]{xcolor}
\usepackage{tikz}
\usepackage{booktabs, tabularx}
\usepackage{rotating, colortbl}
\usepackage{tikz-feynman}

\newcommand{\PreserveBackslash}[1]{\let\temp=\\#1\let\\=\temp}
\newcolumntype{C}[1]{>{\PreserveBackslash\centering}p{#1}}

\newcolumntype{M}{>{\centering\arraybackslash}m{2.25cm}}

\def\thefootnote{\fnsymbol{footnote}} 
\allowdisplaybreaks

\definecolor{darkpink}{RGB}{219, 48, 122}

\begin{document}

\vspace{0.2cm}

\begin{center} 
{\Large\bf A Modular $SU(5)$ Littlest Seesaw}
\end{center}

\begin{center}
{\bf Ivo de Medeiros Varzielas$^1$}\footnote{Email: \tt ivo.de@udo.edu}
,
{\bf Steve F. King$^2$}\footnote{Email: \tt s.f.king@soton.ac.uk}
and 
{\bf Miguel Levy$^1$}\footnote{Email: \tt miguelplevy@ist.utl.pt}
\\\vspace{5mm}
{$^1$CFTP, Departamento de F\'{\i}sica, Instituto Superior T\'{e}cnico,}\\
Universidade de Lisboa,
Avenida Rovisco Pais 1, 1049 Lisboa, Portugal \\
{$^2$ School of Physics and Astronomy, University of Southampton,\\
Southampton SO17 1BJ, United Kingdom } 
\\
\end{center}

\vspace{1.5cm} 

\begin{abstract} 
We extend the littlest modular seesaw to a Grand Unified scenario based on $SU(5)$ endowed with three modular $S_4$ symmetries.  
We leverage symmetry protected zeroes in the leptonic and down quark sectors to suppress deviations to the littlest modular seesaw predictions, but not contributions to the quark mixing.  
The model is supplemented by two weighton fields, such that the hierarchical nature of the charged-lepton masses, as well as the quark masses and mixing, stem from the content and symmetries of the model, rather than a hierarchical nature of the Yukawa coefficients.
\end{abstract}

\begin{flushleft}
\hspace{0.8cm} PACS number(s): 14.60.Pq, 11.30.Hv, 12.60.Fr \\
\hspace{0.8cm} Keywords: Lepton flavour mixing, flavour symmetry
\end{flushleft}

\def\thefootnote{\arabic{footnote}}
\setcounter{footnote}{0}

\newpage

\section{Introduction}

The Standard Model (SM), though successful, does not provide any insight into the origin of fermion families, nor their curious and hierarchical pattern of masses and mixing parameters, which, including also the contrasting neutrino sector, is known as 
the flavour problem. One approach to the flavour problem is based on flavour symmetries, where the approximate tri-bimaximal nature of large solar and atmospheric neutrino mixing, together with smaller reactor mixing, motivates the use of simple non-Abelian discrete symmetries such as $A_4$, $S_4$ and $A_5$~\cite{King:2017guk,Xing:2019vks,Feruglio:2019ktm}. 
Modular invariance~\cite{Ferrara:1989bc,Ferrara:1989qb} can provide the origin of such symmetries in terms of levels $N=3,4,5$,
respectively, of the quotient group of the modular group with its principle congruence subgroup~\cite{Feruglio:2017spp}, leading to very predictive models of flavour. Of special relevance are the fixed points or stabilizers~\cite{Ding:2019gof,deMedeirosVarzielas:2020kji} where for certain values for the modulus, subgroups of the modular transformations are preserved. This has been generalised to the case of multiple modular symmetries in~\cite{deMedeirosVarzielas:2019cyj,King:2019vhv, deMedeirosVarzielas:2021pug, deMedeirosVarzielas:2022ihu}. In order to account for the mass hierarchy of the fermions, an extra singlet field called a weighton~\cite{Criado:2019tzk,King:2020qaj} may be introduced, without the requirement of an extra symmetry.
The origin of all quark and lepton masses and mixing may be addressed by combining Grand unified theories (GUTs)
with modular symmetry groups, for example $SU(5)$ GUT models at level 2~\cite{Kobayashi:2019rzp,Du:2020ylx}, level 3~\cite{deAnda:2018ecu,Chen:2021zty,Charalampous:2021gmf} and level 4~\cite{Zhao:2021jxg,King:2021fhl,Ding:2021zbg}
\footnote{Also flipped $SU(5)\times A_4$~\cite{Charalampous:2021gmf} and $SO(10)\times A_4$ modular models in~\cite{Ding:2021eva} have been considered.}. 

Within the framework of the type Ia seesaw mechanism~\cite{Minkowski:1977sc, Gell-Mann:1979vob, Yanagida:1979as, Glashow:1979nm, Mohapatra:1979ia}, sequential dominance (SD)~\cite{King:1998jw,King:1999cm} of right-handed neutrinos leads to an effective two right-handed neutrino (2RHN) model~\cite{King:1999mb,Frampton:2002qc} with a natural explanation for the physical neutrino mass hierarchy, with normal ordering and the lightest neutrino being approximately massless, $m_1=0$. Constrained sequential dominance (CSD)~\cite{King:2005bj,Antusch:2011ic,King:2013iva,King:2015dvf,King:2016yvg,Ballett:2016yod,King:2018fqh,King:2013xba,King:2013hoa,Bjorkeroth:2014vha} assumes that the two columns of the Dirac neutrino mass matrix are proportional to $(0,1, -1)$ and $(1, n, 2-n)$ respectively, or a related structure,
in the RHN diagonal basis, where the parameter $n$ may be a real number. For example the CSD($\sim 3$) (also called Littlest Seesaw model) is both highly predictive and phenomenologically successful
~\cite{King:2013iva,King:2015dvf,King:2016yvg,Ballett:2016yod,King:2018fqh,King:2013xba,King:2013hoa,Chen:2019oey}.
Remarkably, modular symmetry suggests CSD($1+\sqrt{6}\approx 3.45)$~\cite{Ding:2019gof,Ding:2021zbg}, 
where the three required moduli have been incorporated into complete models of leptons at the field theory level~\cite{deMedeirosVarzielas:2022fbw}, or in 10-dimensional orbifolds~\cite{deAnda:2023udh}. However, it remains to be seen if such models can also accommodate the quark sector non-trivially.

In this paper we consider a $SU(5)$ GUT model with three modular $S_4$ symmetries, which can lead to a predictive Littlest Modular Seesaw model of leptons~\cite{deMedeirosVarzielas:2022fbw}, while at the same time accommodating the quark masses and CKM mixing parameters. In order to address provide a natural explanation of mass and mixing hierarchies, we employ two weighton fields ~\cite{Criado:2019tzk,King:2020qaj}, resulting in a triangular form of hierarchical down-type and quark and charged lepton Yukawa matrices 
as in~\cite{King:2021fhl}. The resulting hierarchical triangular forms preserve the successes of the Littlest Seesaw model while allowing down-type contributions to the CKM angles, with the weightons providing the hierarchical suppressions in all charged fermion sectors, including the up-type quark Yukawa matrix. We present benchmark points which demonstrate the viability of the approach, and show how higher order operators may be controlled by judicious use of the modular weights across all three $S_4$ sectors.

The layout of the remainder of the paper is as follows:
we start by quickly going through the modular framework in Section~\ref{sec:modIntro}, followed by a brief introduction of the $SU(5)$ embedding of the model, shown in Section~\ref{sec:SU5}.  
Following, we present the model in Section~\ref{sec:model}, including the numerical results.  
Finally, we conclude in section~\ref{sec:conc}.

\section{(Multiple) Modular Invariance Framework \label{sec:modIntro}}

The Littlest Modular Seesaw relies on multiple modular symmetries to impose the CSD($n$) structure, with bi-triplet flavons which acquire vacuum expectation values (vevs) in such a way that the three $S_4$ modular symmetries are broken down to a diagonal $S_4$ subgroup which effectively mimics a single modular symmetry, with different moduli, depending on the invariant considered~\cite{deMedeirosVarzielas:2019cyj}.  
The inclusion of flavons (with non-zero vevs) will spontaneously break the modular symmetry, such that it is no longer always possible to perform a modular group action $\gamma$ such that only the fundamental domain may be considered~\cite{Novichkov:2018ovf}.  
In the low-energy theory (when the modular symmetry is broken), the whole domain is relevant, and we can make use of all of the different fixed points~\cite{Ding:2019gof, deMedeirosVarzielas:2020kji}. 
This is also possible to understand in the context of multiple modular symmetries, by  recalling that the bi-triplets break the multiple modular symmetries into a diagonal subgroup.  
As such, the group action will transform all moduli simultaneously, and consequently it is no longer possible to send all moduli to the fundamental domain in general.
In this section we briefly review the modular symmetry framework and the extension to multiple modular symmetries.


The modular group $\Gamma$ is defined by three generators~\cite{Feruglio:2017spp, Novichkov:2020eep}, 
\begin{eqnarray}
S = \begin{pmatrix} 0 & 1 \\ -1 & 0 \end{pmatrix}\, , \quad 
T = \begin{pmatrix} 1 & 1 \\ 1 & 0 \end{pmatrix}\, , \quad 
R = \begin{pmatrix} -1 & 0 \\ 0 & -1 \end{pmatrix}\, ,
\end{eqnarray}
which obey $S^2 = R$, and $(ST)^3 = R^2=\mathbb{1}$, together with $TR = RT$.  
A group element $\gamma$ acts on the modulus $\tau$ (with ${\rm Im}(\tau)>0$) via fractional linear transformations:
\begin{eqnarray}
\gamma = \begin{pmatrix} a & b \\ c & d \end{pmatrix} \in \Gamma \, : \quad \tau \to \gamma \tau = \frac{a \tau + b}{c \tau + d}\,,
\end{eqnarray}
$a, b, c, d$ are integers with $ad-bc=1$. 
By taking these integers as multiples of some integer $N$ we obtain
\begin{eqnarray}
\Gamma(N) = \left\{ \begin{pmatrix} a & b \\ c & d \end{pmatrix} \in PSL(2,\mathbb{Z}), ~~  \begin{pmatrix} a & b \\ c & d \end{pmatrix} = \begin{pmatrix} 1 & 0 \\ 0 & 1 \end{pmatrix} ~~ ({\rm mod}~ N) \right\} \,.
\end{eqnarray}
In practice, the relevant finite group is the quotient of these two infinite groups, $\Gamma_N = \Gamma/\Gamma(N)$, as we consider the representations of fields under $\Gamma_N$.
In terms of the generators, $\Gamma_N$ corresponds to imposing $T^N = \mathbb{1}$.

The chiral superfields $\phi_i$ now transform under $\Gamma_N$ as weighted representations~\cite{Ferrara:1989bc},
\begin{eqnarray}
 \phi_i(\tau) \to \phi_i(\gamma\tau) = (c\tau + d)^{-k_i} \rho_{i j}(\gamma) \phi_j(\tau)\,,
 \label{eq:field_transformation}
\end{eqnarray}
where $k$ is the modular weight, and $\rho$ is a unitary representation of $\Gamma_N$. 

A notable difference of modular symmetries is that the would-be coefficients can be functions of $\tau$ and this then allows for these functions to transform as multiples under $\Gamma_N$ - the modular forms:
\begin{eqnarray} \label{eq:form_transformation}
Y(\tau) \to Y(\gamma \tau) = (c\tau + d)^{k_Y} \rho_{Y}(\gamma) Y(\tau) \,.
\end{eqnarray}

An invariant term of the superpotential is then written as 
\begin{eqnarray}
\big(Y(\tau) \psi_1 \dots \psi_n\big)_\mathbf{1} \, , 
\end{eqnarray}
which is invariant as long as the tensor product of the $\Gamma_N$ representations contains a singlet (\textit{i.e.}, $\rho_Y \otimes \rho_1 \otimes \dots \otimes \rho_n \supset \mathbf{1}$), and the weights cancel out, $k_Y = k_1 + \dots + k_n$, such that the term is unaffected by a modular action of $\gamma$.\footnote{Here, we do not dwell on the choice of normalisations for the modular forms. We assume the canonical renormalisation effect due to the (minimal) Kähler potential (see also \cite{Chen:2019ewa}) to be absorbed into the modular form normalisation.  The relevance of this for the concept of \textit{naturalness} requires a dedicated study~\cite{deMedeirosVarzielas:2023crv, Novichkov:2021evw}.}

Generalizing the framework to include multiple modular symmetries is relatively straightforward in that one adds a modulus $\tau_J$ for each of the modular symmetries $\Gamma_N^{J}$ for $J=1,..., M$ (in general the $N$ can be different for each $J$, but we don't consider it here so we will keep the notation simpler).

A chiral superfield $\phi_i$ now transforms under the multiple (commutative) symmetries as
\begin{eqnarray}
 \phi_i(\tau_1, ...,\tau_M) &\to& \phi_i(\gamma_1\tau_1, ..., \gamma_M \tau_M) \nonumber\\
 &&= \prod_{J=1,...,M} (c_J\tau_J + d_J)^{-k_{i,J}} \bigotimes_{J=1,...,M} \rho_{I_{i,J}}(\gamma_J) \phi_i(\tau_1, \tau_2, ...,\tau_M)\,,
 \label{eq:field_transformation2}
\end{eqnarray}

The modular forms are likewise generalized:
\begin{eqnarray} \label{eq:form_transformation2}
\hspace{-5mm}
&&Y_{(I_{Y,1}, ..., I_{Y,M})}(\tau_1,..., \tau_M) \to Y_{(I_{Y,1},..., I_{Y,M})}(\gamma_1 \tau_1, ..., \gamma_M \tau_M) \nonumber\\
 &&\hspace{2cm}= \prod_{J=1,...,M} (c_J\tau_J + d_J)^{k_{Y,J}}
 \bigotimes_{J=1,...,M} \rho_{I_{Y,J}}(\gamma_J) Y_{(I_{Y,1},..., I_{Y,M})}(\tau_1,..., \tau_M) \,.
\end{eqnarray} 

Another aspect of (multiple) modular symmetry models is that the modular weights can be used to employ the weighton mechanism ~\cite{Criado:2019tzk,King:2020qaj}. By adding a weighton field which transforms typically as a singlet under the modular symmetry, but carries a non-vanishing modular weight, it is possible to arrange terms to be invariant with specific powers of this weighton field, thus suppressing the respective term, in a manner similar to the Froggatt-Nielsen mechanism \cite{Froggatt:1978nt} - but without having to introduce extra symmetry.

\section{$SU(5)$ Embedding \label{sec:SU5} }

We are extending the Littlest Modular Seesaw to a grand unified setting, and a straightforward possibility is to extend the gauge symmetry to a $SU(5)$ framework. 
We briefly review some $SU(5)$ details and set our conventions. Further details about Grand Unified Theories can be found e.g. in \cite{Slansky:1981yr, Ross:1985ai}.
We furnish a $\mathbf{10}$ and a $\mathbf{\overline{5}}$ $SU(5)$ representations with the usual SM fields (including singlet heavy neutrinos) as follows:
\begin{equation}
T=
\begin{pmatrix}
0 & u_G^c & -u_B^c & u_R & d_R \\
& 0 & u_R^c & u_B & d_B \\
& & 0 & u_G & d_G \\
& & & 0 & e^c \\
& & & & 0 
\end{pmatrix} \sim \mathbf{10} \, , \qquad 
F=
\begin{pmatrix}
d_R^c \\
d_B^c \\
d_G^c \\
e^- \\
- \nu
\end{pmatrix} \sim \mathbf{\overline{5}} \, , \quad 
N^c \sim \mathbf{1}\, , 
\end{equation}
 where the $\mathbf{10}$ is an anti-symmetric representation of $SU(5)$, and so we omit the lower entries.  
 
The relevant tensor products for the Yukawa terms are  
\begin{subequations}
\begin{eqnarray}
Y_{\ell}, Y_d \, : &\quad& F \otimes T = \mathbf{\overline{5}} \otimes \mathbf{10} =  \mathbf{5} \oplus \mathbf{45} \, , \\
Y_u \, : &\quad& T \otimes T = \mathbf{10} \otimes \mathbf{10} =  \mathbf{\overline{5}} \oplus \mathbf{\overline{45}} \oplus \mathbf{\overline{50}} \, , \\
Y_D \, : &\quad& F \otimes N = \mathbf{\overline{5}} \otimes \mathbf{1} =  \mathbf{5}  \, , 
\end{eqnarray}
\end{subequations}
where $Y_{u,d}$ are the quark Yukawa matrices, and $Y_\ell$ and $Y_D$ are the charged-lepton and Dirac neutrino mass matrices, respectively.  
We see that we must include scalars in a $\mathbf{\overline{5}}$ representation to have a non-zero $Y_D$, which automatically also leads to a non-zero $Y_\ell$ and $Y_d$.  
A minimal choice which provides a non-zero $Y_u$ is to include a scalar in a $\mathbf{{5}}$ representation.  
As we can see, these gauge assignments (required for the low-energy theory to be SM-like), relate the charged-leptons and down quarks, placing them in a single $SU(5)$ multiplet.  
More specifically, if we include only a $\mathbf{\overline{5}}$ scalar field, responsible for the Yukawas for the (low-energy) charged-leptons and down quarks, we unavoidably find (at the UV scale)
\begin{equation}
Y_d = Y_\ell^T\, .
\end{equation}
This simple relation is not viable, and can be relaxed by the inclusion of a second scalar multiplet. Introducing a $\mathbf{\overline{45}}$, provides a splitting between $Y_\ell$ and $Y_d$~\cite{Georgi:1979df}.  
With the inclusion of these two multiplets, the mass matrices are given by:
\begin{equation}\label{eq:su5dl}
Y_\ell = \left( Y_\mathbf{\overline{5}} - 3Y_\mathbf{\overline{45}} \right) \, , \qquad Y_d = \left( Y_\mathbf{\overline{5}}  + Y_\mathbf{\overline{45}} \right)^T\, , 
\end{equation}
relations which can be inverted to yield 
\begin{equation}
Y_\mathbf{\overline{5}} = \frac{1}{4} \left( Y_\ell + 3 Y_d^T \right) \, , \qquad Y_\mathbf{\overline{45}} = \frac{1}{4} \left( Y_d^T - Y_\ell \right)\, .
\end{equation}
As a consequence, we see that the Yukawa matrices for the down quarks and charged-leptons become general.  
Nonetheless, we also see that texture zeroes are shared by both matrices, up to transposition.  
Thus, even though we include a second scalar multiplet to avoid the stringent relation $Y_d = Y_\ell^T$, the connection between charged-leptons and down quarks lingers on (a situation we will denote as $Y_\ell \sim Y_d^T$ during the rest of the paper).   
A second consequence of the choice of $SU(5)$ as a gauge symmetry comes from the terms responsible for the up quarks Yukawa matrix.  
Since $T$ contains both the LH and RH quarks, the Yukawa terms are given by $T_i H_5 T_j$, and are thus necessarily symmetric.  

In summary, for our purposes here, in order to convert the Littlest Modular Seesaw into $SU(5)$ grand unification, we need to consider non-trivial constraints on the Yukawa couplings.
As had already been mentioned above, we will have symmetrical up quark Yukawa couplings and introduce additional scalars to make the Yukawa couplings of the charged lepton and down quarks viable.  
However, the modular symmetries will lead to texture zeroes in the charged-lepton mass matrices, which are preserved up to transposition, retaining consequences of the $SU(5)$ unification.

\section{The Model \label{sec:model}}

The littlest modular seesaw model is a simple implementation of multiple modular symmetries that economically explains the leptonic sector flavour observables.  
The inclusion of an symmetry based explanation for the quark observables is a desirable next step.  
One interesting possibility is to take advantage of the $SU(5)$ link between charged-leptons and down quarks.  
The inclusion of a $\mathbf{\overline{45}}$ decouples this connection, except that it retains the symmetry protected zeroes in the mass matrices.  
We leverage this fact to design a model in which the symmetries still safeguard the littlest modular seesaw against large contributions from the leptonic sector, while enhancing the contribution of the down sector to the quark mixing.
To retain the successes of the CSD($1\pm\sqrt{6}$) lepton mixing predictions, the structure for $Y_\ell$ cannot have large deviations from the diagonal shape.  
On the other hand, given the connection between the lepton and down-quark sectors ($Y_d \sim Y_\ell^T$), we need to be careful not to suppress the $Y_d$ contribution to the quark mixing.  
To this end, our goal is to employ lower and upper triangular Yukawa matrices for the charged-lepton and down-quark sectors, respectively.  
The point of the triangular shape for $Y_\ell$ is to suppress the corrections to the lepton mixing (compared to the diagonal structure in the unspoiled CSD($n$) framework).  
This suppression can be understood through a simplistic illustration of the $2 \times 2$ case:  
taking the matrix
\begin{equation}
Y = \begin{pmatrix} m_{11} & x \\ 0 & m_{22} \end{pmatrix} \, ,
\end{equation}
we compute the Hermitian matrices (assuming real Yukawas, for simplicity) $H_\ell = Y Y^T$ and $H_d = Y^T Y$.\footnote{In this work, we follow the left-right convention for the Yukawa matrices.}  
The ensuing rotation angles are, assuming $m^2_{22} \gg m^2_{11}$:
\begin{equation}\label{eq:t2a}
\tan\left( 2 \theta_\ell \right) \approx \frac{2 \, x}{m_{22}}\, \frac{m_{11}}{m_{22}} \, , \qquad \tan\left( 2 \theta_d \right) \approx \frac{2 \, x}{m_{22}} \, .
\end{equation}
Clearly, we see that in the lower triangular case, the mixing will be suppressed due to hierarchical nature of the fermion masses, whereas the mixing in the upper triangular case can be $\mathcal{O}(1)$.\footnote{Due to the RH rotation freedom in the SM, we can more accurately describe the lower and upper triangular forms through their hermitian combinations
\begin{equation}
H_\text{lower} = \begin{pmatrix} \lvert m_{11}\rvert ^2 & m_{11} \, x^* \\ m_{11}^* \, x & \lvert m_{22} \rvert^2 + \lvert x \rvert^2 \end{pmatrix} \, , \qquad  
H_\text{upper} = \begin{pmatrix} \lvert m_{11}\rvert ^2 + \lvert x \rvert ^2 & m_{22} \, x^* \\ m_{22}^* \, x & \lvert m_{22} \rvert^2 \end{pmatrix} \, , 
\end{equation} 
together with $m_{22} \gg m_{11}$, rather than their unphysical Yukawa shapes.  
Regardless, we feel no confusion will arise throughout the paper.  
}  

The UV nature of the model requires the running of the measured masses to some high scale.   
Here, we make use of the values shown in~\cite{Okada:2020rjb}, obtained from~\cite{Antusch:2013jca, Bjorkeroth:2015ora}, for $\tan\beta=5$ and a GUT scale at \mbox{$2 \times 10^{16}$ GeV}:\footnote{We consider for simplicity that $SU(5)$ is broken at \mbox{$2 \times 10^{16}$ GeV}, and use the SM field content for calculating the running.} 
\begin{subequations}
\begin{eqnarray}
\begin{matrix}
y_e = ( 1.97 \pm 0.024) \times 10^{-6}\, , \qquad &y_\mu = ( 4.16 \pm 0.050) \times 10^{-4}\, , \qquad &y_\tau = (7.07 \pm 0.073) \times 10^{-3} \, ,  \\
y_u = (2.92 \pm 1.81) \times 10^{-6}\, , \qquad &y_c = (1.43 \pm 0.100) \times 10^{-3}\, , \qquad &y_t = (0.534 \pm 0.0341) \times 10^{0}\, , \\
y_d = (4.81 \pm 1.06) \times 10^{-6}\, , \qquad &y_s = (9.52\pm 1.03) \times 10^{-5}\, , \qquad &y_b = (6.95 \pm 0.175) \times 10^{-3}\, , \\
\theta_{12} = (13.027 \pm 0.0814)\,^{\circ} \, , \qquad &\theta_{23} = (2.054\pm 0.384)\,^{\circ} \, , \qquad &\theta_{13} = (0.1802 \pm 0.0281)\,^{\circ}\, ,
\end{matrix}
\end{eqnarray}
together with
\begin{equation}
\delta  = ( 69.21 \pm 6.19)\,^{\circ}.
\end{equation}
\end{subequations}
For the neutrino observables, we use the NuFit~5.2 IR values~\cite{Esteban:2020cvm, nufit}.\footnote{We assume the neutrino observables have negligible running, as done in~\cite{Bjorkeroth:2015ora}. See also Ref.~\cite{Geib:2017bsw} for a comprehensive analysis.}

Taking the Cabibbo angle as a measure, $\lambda \sim 0.227$, the experimental values are approximately
\begin{eqnarray}\label{eq:suppressions}
\begin{matrix}
y_e \sim \lambda^{8.9}\, , \qquad & y_\mu \sim \lambda^{5.3}\, , \qquad & y_\tau \sim \lambda^{3.3}\, , \\
y_u \sim \lambda^{8.6}\, , \qquad &y_c \sim \lambda^{4.4}\, , \qquad &y_t \sim \lambda^{0.4}\, , \\
y_d \sim \lambda^{8.2}\, , \qquad &y_s \sim \lambda^{6.2}\, , \qquad &y_b \sim \lambda^{3.4}\, , \\
\theta_{12} \sim \lambda^1 \, , \qquad &\theta_{23} \sim \lambda^{2.3} \, , \qquad &\theta_{13} \sim \lambda^{3.9}.  
\end{matrix}
\end{eqnarray}
We wish to have $\mathcal{O}(1)$ coefficients controlling the fermionic masses and mixings. Consequently, we can exploit the smallness of the quark mixing angles and naively use Eq.~\eqref{eq:t2a} to populate the entries of the Yukawa matrices, such that these are, by design, suppressed in such a way that both the quark mass hierarchy and the CKM matrix come out naturally:
\begin{eqnarray}\label{eq:goal}
Y_u \sim \begin{pmatrix} \lambda^8 & \lambda^5 & \lambda^4 \\ \lambda^5 & \lambda^4 & \lambda^2 \\ \lambda^4 & \lambda^2 & \lambda^0 \end{pmatrix} \, , \qquad
Y_d^T \sim \begin{pmatrix} \lambda^8 & 0 & 0 \\ \lambda^7 & \lambda^6 &  0 \\ \lambda^7 & \lambda^5 & \lambda^3 \end{pmatrix} \, , \qquad 
Y_\ell \sim \begin{pmatrix} \lambda^9 & 0 & 0 \\ - & \lambda^5 & 0 \\ - & - & \lambda^3 \end{pmatrix} \, , 
\end{eqnarray}
where the off-diagonal non-zero entries of $Y_\ell$ are undetermined and denoted as ``$-$'', since the lower triangular shape suppresses the contributions to the leptonic mixing, and the main driver behind the PMNS mixing matrix comes from the Dirac neutrino structure, in the modular CSD($1\pm\sqrt{6}$) set-up.  
As for the neutrino sector, due to the built-in suppression mechanism in the form of the Type-I seesaw, we do not require any specific suppressions.  
We stress that the matrices shown in Eq.~\eqref{eq:goal} are derived merely from the experimental values, and are not necessarily attainable in a specific set-up.  
Indeed, the $SU(5)$ gauge symmetry, assuming the same order of magnitude for all Yukawas, will forbid different suppressions in $Y_\ell$ and $Y_d^T$, as we will see later.

Our model relies on a $SU(5)$ gauge symmetry, supplemented by 3 distinct $S_4$ modular symmetries.  
The assignments of the fields are given in Table~\ref{tab:model}, both for the gauge and modular symmetries.  
We note that although we use non-integer weights for the fields, only even-weighted Yukawa modular forms are considered, consistent with the requirement of invariance under the $S_4$ modular group.  
Indeed, rational modular weights for the fields are also obtained from top-down constructions in~\cite{Nilles:2020kgo, Baur:2021bly, Baur:2022hma}.  
As such, the present framework continues to be that of modular invariance, and not that of metaplectic models~\cite{Liu:2020msy, Yao:2020zml}, since we do not consider half-integer modular forms.

\begin{table}[h!]
\centering
\renewcommand{\arraystretch}{1.2}
\begin{tabular}{|c| c | c c | c c | c c |}
\hline
Field 				& 	$SU(5)$	 				&	${S}_4^A$ 		&	$k_A$ 	&	${S}_4^B$ 		&	$k_B$ 		&	${S}_4^C$ 		& 	$k_C$ 		\\
\hline \hline
$F$ 					&	$\mathbf{\overline{5}}$				& 	$\mathbf{1}$		&	$+\frac{1}{2}$	&	 $\mathbf{1}$		&	$+\frac{1}{2}$	&	$\mathbf{3}$		&	$-3$		\\
$T_1$				&	$\mathbf{10}$			& 	$\mathbf{1}$		&	$+1$\		&	 $\mathbf{1}$		&	$+1$		&	$\mathbf{1'}$		&	$+3$		\\
$T_2$				&	$\mathbf{10}$			& 	$\mathbf{1}$		&	$+\frac{1}{2}$	&	 $\mathbf{1}$		&	$+\frac{1}{2}$	&	$\mathbf{1'}$		&	$+3$		\\
$T_3$				&	$\mathbf{10}$			& 	$\mathbf{1}$		&	$  0$		&	 $\mathbf{1}$		&	$  0$		&	$\mathbf{1'}$		&	$+3$		\\
$N_A^c$				&	$\mathbf{1}$				& 	$\mathbf{1'}$		&	$+\frac{9}{2}$	& 	$\mathbf{1}$		&	$+\frac{1}{2}$	& 	$\mathbf{1}$		&	$-1$			\\
$N_B^c$				&	$\mathbf{1}$				& 	$\mathbf{1}$		&	$+\frac{1}{2}$	& 	$\mathbf{1'}$		&	$+\frac{5}{2}$	& 	$\mathbf{1}$		&	$-1$			\\
\hline
$\Phi_{AC}$			&	$\mathbf{1}$				&	$\mathbf{3}$		&	$  0$		&	$\mathbf{1}$		&	$  0$		&	$\mathbf{3}$		&	$  0$		\\
$\Phi_{BC}$			&	$\mathbf{1}$				&	$\mathbf{1}$		&	$  0$		&	$\mathbf{3}$		&	$  0$		&	$\mathbf{3}$		&	$  0$		\\
\hline
$\phi_{T}$			&	$\mathbf{1}$				&	$\mathbf{1}$		&	$-\frac{1}{2}$	&	$\mathbf{1}$		&	$-\frac{1}{2}$	&	$\mathbf{1}$		&	$  0$		\\
$\phi_{F}$			&	$\mathbf{1}$				&	$\mathbf{1}$		&	$-\frac{1}{2}$	&	$\mathbf{1}$		&	$-\frac{1}{2}$	&	$\mathbf{1}$		&	$+2$		\\
\hline
$H_5$				&	$\mathbf{5}$				&	$\mathbf{1}$		&	$ 0$			&	$\mathbf{1}$		&	$ 0$			&	$\mathbf{1}$		&	$ 0$			\\
$H_{\overline{5}}$		&  $\mathbf{\overline{5}}$		&	$\mathbf{1}$		&	$ 0$			&	$\mathbf{1}$		&	$ 0$			&	$\mathbf{1}$		&	$ 0$			\\
$H_{\overline{45}}$	&  $\mathbf{\overline{45}}$		&	$\mathbf{1}$		&	$ 0$			&	$\mathbf{1}$		&	$ 0$			&	$\mathbf{1}$		&	$ 0$			\\
\hline
\end{tabular}
\caption{\label{tab:model} Assignments of the fields under the $SU(5)$ gauge symmetry and the representations and weights under the 3 modular symmetries ($S_4^A$, $S_4^B$, $S_4^C$) considered.  
We omit fields that are necessary for a consistent UV completion, such as messenger fields to complete the non-renormalizable terms, as well as the driving fields responsible for the bi-triplet and weighton VEVs.}
\end{table}

Motivated by the modular CSD($1\pm\sqrt{6}$) structure, we do not take the values for the moduli as free parameters, and set them to the relevant stabilisers~\cite{Ding:2019gof, deMedeirosVarzielas:2020kji}.  
As such, we take 
\begin{subequations}\label{eq:stabs}
\begin{eqnarray}
&&\tau_A = \frac{1}{2} + \frac{i}{2} \, , \qquad  \tau_B = \textcolor{white}{-}\frac{3}{2} + \frac{i}{2} \, , \qquad \tau_C = \omega \, , \label{eq:Plus}\\
&&\tau_A = \frac{1}{2} + \frac{i}{2} \, , \qquad \tau'_B = -\frac{1}{2} + \frac{i}{2}  \, , \qquad \tau_C = \omega \, ,\label{eq:Minus}
\end{eqnarray}
where the choice of Eq.~\eqref{eq:Plus} corresponds to the CSD($1 + \sqrt{6}$) case, whereas Eq.~\eqref{eq:Minus} gives rise to CSD($1 - \sqrt{6}$).  
In the basis of~\cite{Ding:2019gof, Ding:2021zbg}, these correspond to more familiar fixed points:
\begin{eqnarray}
&&\tau_A = 2+i \, , \qquad  \tau_B = i \, , \qquad \qquad \qquad \tau_C = \omega \, , \label{eq:Plusnew}\\
&&\tau_A = 2+i \, , \qquad \tau'_B = -\frac{8}{13} + \frac{i}{13}  \, , \qquad \tau_C = \omega \, .
\end{eqnarray}
\end{subequations}
We emphasise that the fixed points in Eq.\ref{eq:Plusnew} arise from the well known single modulus fixed points $\tau = i$ and $\tau = \omega$. With three moduli, the two fixed points $i$ and $i+2$ are simply related but inequivalent.

The superpotential responsible for the up-quark mass matrix comes from $T_i H_5 T_j$ couplings:
\begin{eqnarray}\label{eq:wu}
&w_u =H_5\bigg\{ 
 T_1 \left[ y_{uu} Y_\mathbf{1}^{(6)}\left(\tau_C\right)  \left(\dfrac{\phi_T^4}{\Lambda^4}\right)  
 +  {y}'_{uu}  Y_\mathbf{1}^{(12)}\left(\tau_C\right) \left(\dfrac{\phi_F^3 \phi_T}{\Lambda^4}\right) \right] T_1  + 
 T_2 \left[  y_{cc} Y_\mathbf{1}^{(6)}\left(\tau_C\right)  \left( \dfrac{\phi_T^2}{\Lambda^2}\right) \right] T_2   \nonumber & \\
&
+ y_{tt} Y_\mathbf{1}^{(6)}\left(\tau_C\right) \left[T_3 T_3\right]
 + T_1 \left[ y_{uc} Y_\mathbf{1}^{(6)}\left(\tau_C\right) \left(\dfrac{\phi_T^3}{\Lambda^3}\right)  
 +  {y}'_{uc}Y_\mathbf{1}^{(12)}\left(\tau_C\right) \left( \dfrac{\phi_F^3}{\Lambda^3}\right) \right] T_2  
  \nonumber & \\
&
 + T_1 \left[ y_{ut} Y_\mathbf{1}^{(6)}\left(\tau_C\right)  \left(\dfrac{\phi_T^2}{\Lambda^2}\right) \right] T_3 + 
 T_2 \left[ y_{ct} Y_\mathbf{1}^{(6)}\left(\tau_C\right) \left(\dfrac{\phi_T}{\Lambda}\right) \right] T_3 
 \bigg\}  , 
\end{eqnarray}
where we suppress $Y_\mathbf{1}^{(0)}\left(\tau_{A,B}\right)$ from the notation, and $\Lambda$ stands for the relevant UV scale for a particular non-renormalizable operator, which we take to be universal for notational convenience.    
The contributions to $Y_d$ (and similarly for $Y_\ell$) come from the couplings to both $H_{\overline{5}}$ and $H_{\overline{45}}$.  
The superpotential will read the same for both scalars, with $H_{\overline{5}}$ and $H_{\overline{45}}$ exchanged, and different Yukawa couplings:
\begin{eqnarray}\label{eq:wd}
& w_{\ell, d} = H_{\overline{5}} \bigg\{
F \left[ y_{11}^{\overline{\mathbf{5}}} Y_\mathbf{3'}^{(6)}\left(\tau_C\right) \left(\dfrac{\phi_F^3}{\Lambda^3}\right) + 
 y_{12}^{\overline{\mathbf{5}}} Y_\mathbf{3}^{(4)}\left(\tau_C\right) \left(\dfrac{\phi_F^2 \phi_T}{\Lambda^3}\right) + 
 y_{13}^{\overline{\mathbf{5}}} Y_\mathbf{3'}^{(2)}\left(\tau_C\right) \left(\dfrac{\phi_F \phi_T^2}{\Lambda^2}\right) \right] T_1  \nonumber & \\
& + \, F \left[ y_{22}^{\overline{\mathbf{5}}} Y_\mathbf{3}^{(4)}\left(\tau_C\right) \left(\dfrac{\phi_F^2}{\Lambda^2}\right) + 
 y_{23}^{\overline{\mathbf{5}}} Y_\mathbf{3'}^{(2)}\left(\tau_C\right)  \left( \dfrac{\phi_F \phi_T}{\Lambda^2}\right) \right] T_2 + 
 F \left[ y_{33}^{\overline{\mathbf{5}}} Y_\mathbf{3'}^{(2)} \left(\tau_C\right) \left(\dfrac{\phi_F}{\Lambda}\right) \right] T_3 \bigg\} &\\
 & + \quad \mathbf{\left(\overline{5} \to \overline{45}\right)}  \nonumber \, . &
\end{eqnarray}
The relevant modular forms (for $\tau_C=\omega$) are given by
\begin{equation}
Y_{\mathbf{3'}}^{(2)}(\tau_C) = \begin{pmatrix} 0 \\ 1 \\ 0 \end{pmatrix} , \quad 
Y_{\mathbf{3}, \mathbf{3'}}^{(4)}(\tau_C) = \begin{pmatrix} 0 \\ 0 \\ 1 \end{pmatrix} , \quad 
Y_{\mathbf{3},\mathbf{3'}}^{(6)}(\tau_C)  = \begin{pmatrix} 1 \\ 0 \\ 0 \end{pmatrix} , \quad 
Y_{\mathbf{1}}^{(6)}(\tau_C) = Y_{\mathbf{1}}^{(12)}(\tau_C)  = 1 \, , 
\end{equation}
and thus, the resulting Yukawa matrices are 
\begin{subequations}
\begin{eqnarray}\label{eq:MassMats}
&Y_u =  
\begin{pmatrix} 
y_{uu} \epsilon_T^4 + y_{uu}' \epsilon_F^3 \epsilon_T & y_{uc} \epsilon_T^3 + y_{uc}' \epsilon_F^3 & y_{ut} \epsilon_T^2 \\
. & y_{cc} \epsilon_T^2 & y_{ct} \epsilon_T \\
.& . & y_{tt} 
\end{pmatrix}\, , & \\
&Y_d = 
\begin{pmatrix}
y_{dd} \epsilon_F^3 & y_{ds} \epsilon_F^2 \epsilon_T & y_{db} \epsilon_F \epsilon_T^2 \\
0 & y_{ss} \epsilon_F^2 & y_{sb} \epsilon_F \epsilon_T \\
0 & 0 & y_{bb} \epsilon_F
\end{pmatrix} \, \qquad 
Y_\ell = 
\begin{pmatrix}
y_{ee} \epsilon_F^3 & 0 & 0 \\
y_{\mu e} \epsilon_F^2 \epsilon_T & y_{\mu \mu} \epsilon_F^2 & 0 \\
y_{\tau e} \epsilon_F \epsilon_T^2 & y_{\tau \mu} \epsilon_F \epsilon_T & y_{\tau \tau} \epsilon_F
\end{pmatrix} \, .
&
\end{eqnarray}
\end{subequations}

The superpotentials of Eqs.~\eqref{eq:wu}~and~\eqref{eq:wd} responsible for the quark and charged-lepton masses rely on non-renormalizable operators, through multiple weighton insertions.  
After the weighton fields acquire a non-zero VEV, $\epsilon_{F, T} = \left< \phi_{F,T} \right>/\Lambda$, the quarks and charged-leptons get contributions to their masses \textit{a la} Froggatt-Nielsen, as per the weighton mechanism.  
If we assume $\mathcal{O}(1)$ coefficients, together with $\epsilon_F \sim \lambda^3$, and $\epsilon_T \sim \lambda^2$, we see that the mass matrices are close to those of Eq.~\eqref{eq:goal}:
\begin{equation}\label{eq:got}
Y_u \sim \begin{pmatrix} \epsilon_T^4 + \epsilon_F^3 \epsilon_T & \epsilon_T^3 + \epsilon_F^3 & \epsilon_T^2 \\ . & \epsilon_T^2 & \epsilon_T \\ . &. & 1 \end{pmatrix},  \qquad
Y_d \sim \begin{pmatrix} \epsilon_F^3 & \epsilon_F^2 \epsilon_T & \epsilon_F \epsilon_T^2 \\ 0 & \epsilon_F^2 & \epsilon_F\epsilon_T \\ 0 & 0 & \epsilon_F \end{pmatrix},  \qquad
Y_\ell \sim \begin{pmatrix} \epsilon_F^3 & 0 & 0 \\ \epsilon_F^2 \epsilon_T  & \epsilon_F^2 & 0 \\ \epsilon_F \epsilon_T^2 & \epsilon_F \epsilon_T & \epsilon_F \end{pmatrix}.
\end{equation}
where the up quark Yukawa matrix is symmetric.
We see that we are unable to get different suppressions for $Y_\ell$ and $Y_d$, due to the $SU(5)$ nature of the model.  
Nonetheless, if we identify $\epsilon_T \sim \lambda^2$ and $\epsilon_F \sim \lambda^3$, all the entries would have the desired suppressions, except for the $(1,2)$ entries of both $Y_u$ and $Y_d$ carry an extra suppression of $\lambda$, and so does the $(1,1)$ entry of $Y_d$ (yielding, however, the correct suppression for $Y_\ell$).  
As such, we expect from the start that the model is able to fit the quark masses and mixings, as well as the charged-lepton masses, with $\mathcal{O}(1)$ coefficients, as the model is designed such that the weighton insertions could be responsible for most of the observed hierarchies.

We turn now to the neutrino sector.  
The neutrino Dirac Yukawa matrix comes from $F H_5 N^c$ couplings.  
Due to the requirement of invariance under the modular symmetries, we see that $Y_D$ can only be non-zero via couplings to the bi-triplets $\Phi_{AC}$ and $\Phi_{BC}$.  
The allowed superpotential for the neutrino Dirac Yukawa matrix reads
\begin{equation}\label{eq:wD}
w_D = H_5 \bigg\{  a\,  Y_\mathbf{3'}^{(4)}\left(\tau_A\right) F \left( \dfrac{\phi_F^2}{\Lambda^2} \dfrac{\left< \Phi_{AC}\right>}{\Lambda}\right) N_A^c + b \, Y_\mathbf{3'}^{(2)}\left(\tau_B\right) F \left( \dfrac{\phi_F^2}{\Lambda^2} \dfrac{\left<\Phi_{BC}\right>}{\Lambda}\right) N_B^c \bigg\} \, .
\end{equation}
Given the moduli of Eq.~\eqref{eq:stabs}, the relevant Yukawa modular forms are
\begin{eqnarray}
Y_\mathbf{3'}^{(4)}\left(\tau_A\right) = \begin{pmatrix} 0 \\ -1 \\ 1 \end{pmatrix} \, , 
\qquad  Y_\mathbf{3'}^{(2)}\left(\tau_B\right) = \begin{pmatrix} 1 \\ 1- \sqrt{6} \\ 1+ \sqrt{6} \end{pmatrix} \, , 
\qquad  Y_\mathbf{3'}^{(2)}\left(\tau'_B\right) = \begin{pmatrix} 1 \\ 1+ \sqrt{6} \\ 1- \sqrt{6} \end{pmatrix} \, .  
\end{eqnarray}
After both the bi-triplets and the weighton acquire a non-zero VEV, the terms of Eq.~\eqref{eq:wD} populate the neutrino Dirac Yukawa matrix as
\begin{equation}
Y_D \propto \epsilon_F^2
\begin{pmatrix}
0 & b \\
a & b\left(1 \pm \sqrt{6}\right) \\
-a & b\left(1 \mp \sqrt{6}\right) \\
\end{pmatrix} \, .
\end{equation}

Lastly, we analyse the relevant terms for the RH neutrino mass matrix.  
The modular assignments of Table~\ref{tab:model} do not allow for the presence of bare mass terms, otherwise allowed by gauge invariance.  
However, similarly for the remaining fermions, we can build non-renormalizable terms which will generate mass terms for the heavy neutrinos below an appropriately large scale.  
The relevant Yukawa modular forms here will transform as $\mathbf{1^{(')}}$ and, as it happened for the cusp, there are vanishing modular forms at $\tau_A$ and $\tau_B$.  
We find that, up to weight 10, the relevant non-zero modular forms are
\begin{equation}
Y_\mathbf{1		}^{(4)} \, , \quad 
Y_\mathbf{1'		}^{(6)} \, , \quad 
Y_\mathbf{1		}^{(8)} \, , \quad 
Y_\mathbf{1'		}^{(10)} \,  \, . 
\end{equation}
The Majorana superpotential is given by
\begin{equation}\label{eq:wM}
w_M = \frac{1}{2} M_A Y_\mathbf{1}^{(8)}\left(\tau_A\right) \left( \dfrac{\phi_F \phi_T}{\Lambda^2}\right) N_A^c N_A^c + \frac{1}{2} M_B Y_\mathbf{1}^{(4)}\left(\tau_B\right) \left( \dfrac{\phi_F \phi_T}{\Lambda^2}\right)  N_B^c N_B^c \, ,
\end{equation}
where, as usual, we omit any modular form of weight 0.  
Any mixed term is forbidden by the modular symmetries.  
The ensuing Majorana mass matrix is then given by
\begin{equation}\label{eq:mM}
M_M = \epsilon_F \epsilon_T \begin{pmatrix} M_A & 0 \\ 0 & M_B \end{pmatrix}\, .
\end{equation}
Through the type-I seesaw mechanism, the effective mass matrix for the light neutrinos becomes
\begin{equation}
m_\nu = M_D \cdot M_R^{-1} \cdot M_D^T = \dfrac{v_u^2 \epsilon_F^3}{\epsilon_T} 
\begingroup
\setlength\arraycolsep{15pt}
\begin{pmatrix}  
\dfrac{b^2}{M_B} & \dfrac{b^2 n}{M_B} &  \dfrac{b^2(2-n)}{M_B} \\[12pt]
. & \dfrac{a^2}{M_A} + \dfrac{b^2 n^2}{M_B} & -\dfrac{a^2}{M_A} + \dfrac{b^2n(2-n)}{M_B} \\[12pt] 
. & . & \dfrac{a^2}{M_A} + \dfrac{b^2(2-n)^2}{M_B}
\end{pmatrix},
\endgroup
\label{eq:mnu_mee}
\end{equation}
with $n=1\pm\sqrt{6}$, as desired for the CSD($1 \pm \sqrt{6}$) predictions.   
This can be written in more compact notation as 
\begin{equation}
m_\nu =m_a
\begin{pmatrix} 0 & 0 & 0 \\ 0 & 1 & -1 \\ 0 & -1 & 1 \end{pmatrix} 
+ m_be^{i \beta} 
\begin{pmatrix} 1 & n & 2-n \\ n & n^2 &  n (2-n) \\ 2-n & n (2-n) & (2-n)^2 \end{pmatrix}.
\end{equation}
where $m_a=\left\|\dfrac{v_u^2 \epsilon_F^3}{\epsilon_T}  \dfrac{a^2}{M_A}\right\|$
and $m_b=\left\|  \dfrac{v_u^2 \epsilon_F^3}{\epsilon_T}    \right\|$ are real parameters and $\beta$ is an undetermined phase.
This shows that the neutrino mass matrix is completely determined by three real parameters $m_a$, $m_b$ and $\beta$,
where $n=1\pm\sqrt{6}$, making it a highly predictive scheme, which successfully describes the current data as recently discussed~\cite{Costa:2023bxw}.

We note as a final remark that, due to the choice of assignments of Table~\ref{tab:model}, the superpotentials of Eqs.~\eqref{eq:wu},~\eqref{eq:wd},~\eqref{eq:wD},~and~\eqref{eq:wM} do not have higher order corrections stemming for further insertions of weightons.  
A more technical note is shown in Appendix~\ref{app:ExtraTerms}, where we highlight the importance of having only one way to generate the couplings of Eq~\eqref{eq:wD}.

We expect the model to be compatible with experiment, due to its design.  
For completeness, we show here one point to showcase that indeed we can get the UV values for the quark masses and mixings with $\mathcal{O}(1)$ coefficients, and negligible $\chi^2$:
\begin{eqnarray}\label{eq:numres}
& 
y_{uu}  = 1.1533 \,e^{- 0.524 i} , \quad
y'_{uu} = 1.0001 \,e^{-2.24 i}, \quad
y_{uc} = 0.97294 \,e^{-2.59 i}, 
& \nonumber \\
&
y'_{uc} = 0.93204 \,e^{0.0393 i }, \quad
y_{ut} =0.97272 \,e^{1.20 i}, \quad
y_{cc} = 1.0264 \,e^{1.53 i}, 
& \nonumber \\
&
y_{ct} = 0.92436 \,e^{ -0.461 i}, \quad
y_{tt} = 0.53034 \,e^{-2.34 i}, 
& \nonumber \\ 
\\
&
y_{dd} = 2.6549 \,e^{1.10i}, \quad
y_{ds} = 2.1282 \,e^{0.555i}, \quad
y_{db} = 1.0022 \,e^{-1.07i}, 
& \nonumber \\
&
y_{ss} = 0.62888 \,e^{-2.97i}, \quad
y_{sb} = 0.93386 \,e^{2.76i}, \quad
y_{bb} = 0.56589 \,e^{1.10 i}, 
& \nonumber \\ 
\nonumber \\
&
\epsilon_T = 4.877 \times 10^{-2} \approx 0.946 \lambda^2, \quad
\epsilon_F = 1.224 \times 10^{-2} \approx 1.046 \lambda^3 \, .
&\nonumber 
\end{eqnarray}

As expected, we see that $\epsilon_F \sim \lambda^3$, and $\epsilon_T \sim \lambda^2$.  
Moreover, we see that we can easily fit the quark masses and mixings with $\mathcal{O}(1)$ Yukawas, due to having the correct suppressions on the Yukawa matrices.  
As mentioned earlier, the $(1,1)$ and $(1,2)$ entries of $Y_d$ have an extra $\lambda$ suppression comparing to our naïve guess.  
In that sense, the associated couplings ($y_{dd}$ and $y_{ds}$) need to be larger to compensate.  
That is clearly seen in the numerical result of Eq.~\eqref{eq:numres}, in the relative hierarchy between $y_{dd, ds}$ and the remaining Yukawas.   
Turning now to the neutrino and charged-leptons, we provide a benchmark for using the results obtained for the quarks, namely, taking $\epsilon_T$ and $\epsilon_F$ from Eq.~\eqref{eq:numres}.  
We find, for the $n=1+\sqrt{6}$ case, and taking into account the SK-atmospheric data:
\begin{eqnarray}
&y_{ee} = 1.0740 e^{2.48 i} , \quad 
y_{\mu e} = 1.0000e^{2.62 i}, \quad
y_{\tau e} = 1.0000 e^{0.495i}, & \nonumber \\
& y_{\mu \mu} = 2.9497 e^{-0.389 i} , \quad
y_{\tau \mu} = 4.0000 e^{0.490 i},  \quad
y_{\tau \tau} = 0.54344 e^{1.31 i} , & \\
& r = 7.317 \times 10^{-2}, \qquad \beta = 1.2378 & \nonumber
\end{eqnarray}
which is very close to the LMS best-fit point, with $\chi^2 \sim 1.9$. 

This good fit to the flavour observables in a $SU(5)$ unified model is obtained through a combination of Georgi-Jarlskog factors, the upper / lower diagonal form for the matrices of the down quarks and charged leptons respectively, and the weighton mechanism, and demonstrate clearly some of the advantages of employing (multiple) modular symmetries in theories of flavour.

\section{Conclusion \label{sec:conc}}

A grand unified theory of flavour is a desirable goal, but not easy to achieve. The connections between families imposed by unification restrict the solutions to the flavour problem. In this paper we tackled the challenge by embedding into $SU(5)$ unification the littlest modular seesaw model. This is a model that describes the lepton sector extremely well through multiple modular flavour symmetries. We surmounted the typical difficulty arising from the relation between charged leptons and down quarks by employing Georgi-Jarlskog factors arising from appropriate $SU(5)$ multiplets, and, from the modular flavour symmetry, constructing an upper-diagonal matrix in the flavour symmetry basis for the down quarks. Due to this, the contributions to quark mixing are sizeable at the same time that the transposed charged lepton matrix is lower-diagonal, such that its off-diagonal entry contributions to the leptonic mixing are suppressed by the hierarchical charged lepton mass ratios.
We employ also two weightons, which enable us to justify the hierarchical entries in the mass matrices through the use of the modular wieghts, a mechanism which is reminiscent of typical Froggatt-Nielsen, but without requiring the introduction of an extra symmetry.
In conclusion, we present an elegant $SU(5)$ unified, multiple modular flavour symmetry model which accounts for the flavour observables in a unified setting, with an emphasis on the predictivity of the leptonic sector, and the use of weightons that explain all the fermion mass hierarchies.

\section*{Acknowledgements}
This work was partially supported by Fundação para a Ciência e a Tecnologia (FCT, Portugal) by the projects~CERN/FIS-PAR/0002/2021,~CERN/FIS-PAR/0019/2021, CFTP-FCT Unit~UIDB/00777/2020 and~UIDP/00777/2020, which are partially funded through POCTI (FEDER), COMPETE, QREN and EU.
IdMV~acknowledges support by FCT fellowships in the framework of the project UIDP/00777/2020. 
M.L.~acknowledges support from FCT through the grant No.~PD/BD/150488/2019, in the framework of the Doctoral Programme IDPASC-PT.
SFK acknowledges the STFC Consolidated Grant ST/L000296/1 and the European Union's Horizon 2020 Research and Innovation programme under Marie Sklodowska-Curie grant agreement HIDDeN European ITN project (H2020-MSCA-ITN-2019//860881-HIDDeN).

\appendix

\section{$S_4$ Group Theory and Modular Forms at the Cusp} \label{app:S4}

The generators of $S_4$ obey 
\begin{eqnarray}
S^2 = (S T)^3 = T^4 = \mathbb{1} \, .
\end{eqnarray}
We follow the $S_4$ basis of~\cite{Novichkov:2018ovf}, where the representation matrices are
\begin{subequations}
\begin{eqnarray}
&\mathbf{1}&  \, : \quad \rho(S) = 1 \, , \quad \rho(T) = 1 \, , \\
&\mathbf{1}'&  \, : \quad \rho(S) = -1 \, , \quad \rho(T) = -1 \, , \\
&\mathbf{2}&  \, : \quad \rho(S) = \begin{pmatrix} 0 & \omega \\ \omega^2 & 0 \end{pmatrix}  \, , \quad \rho(T) =  \begin{pmatrix} 0 & 1 \\ 1 & 0 \end{pmatrix}  \, , \\
&\mathbf{3}&  \, : \quad \rho(S) = \frac{1}{3}
\begin{pmatrix}
 -1 & 2 \omega^2 & 2 \omega \\ 
  2 \omega & 2  & - \omega^2  \\ 
   2 \omega^2 & -\omega & 2 
\end{pmatrix}  \, , \quad 
\rho(T) =  \frac{1}{3}
\begin{pmatrix}
 -1 & 2 \omega & 2 \omega^2 \\ 
  2 \omega & 2 \omega^2  & - 1  \\ 
   2 \omega^2 & -1 & 2 \omega 
\end{pmatrix}  \, , \\
&\mathbf{3}'&  \, : \quad \rho(S) = -\frac{1}{3}
\begin{pmatrix}
 -1 & 2 \omega^2 & 2 \omega \\ 
  2 \omega & 2  & - \omega^2  \\ 
   2 \omega^2 & -\omega & 2 
\end{pmatrix}  \, , \quad 
\rho(T) =  -\frac{1}{3}
\begin{pmatrix}
 -1 & 2 \omega & 2 \omega^2 \\ 
  2 \omega & 2 \omega^2  & - 1  \\ 
   2 \omega^2 & -1 & 2 \omega 
\end{pmatrix}  \, .
\end{eqnarray}
\end{subequations}
The tensor products are given by
\begin{subequations}
\begin{eqnarray}
\mathbf{1} \otimes \mathbf{r} &=& \mathbf{r} \, , \\
\mathbf{1}' \otimes \mathbf{1}' &=& \mathbf{1} \, , \\
\mathbf{1}' \otimes \mathbf{2} &=& \mathbf{2} \, , \\
\mathbf{1}' \otimes \mathbf{3} &=& \mathbf{3}' \, , \\
\mathbf{1}' \otimes \mathbf{3}' &=& \mathbf{3} \, , \\
\mathbf{2} \otimes \mathbf{2} &=& \mathbf{1} \oplus \mathbf{1}' \oplus \mathbf{2} \, , \\
\mathbf{2} \otimes \mathbf{3} &=& \mathbf{3} \oplus \mathbf{3}' \, , \\
\mathbf{2} \otimes \mathbf{3}' &=& \mathbf{3} \oplus \mathbf{3}' \, , \\
\mathbf{3} \otimes \mathbf{3} &=& \mathbf{1} \oplus \mathbf{2} \oplus \mathbf{3} \oplus \mathbf{3}' \, , \\
\mathbf{3} \otimes \mathbf{3}' &=& \mathbf{1}' \oplus \mathbf{2} \oplus \mathbf{3} \oplus \mathbf{3}' \, , 
\end{eqnarray}
\end{subequations}
where we only show the relevant Clebsch-Gordan coefficients for our model, and the remaining can be found in~\cite{Novichkov:2018ovf}:
\begin{eqnarray}
\left(\mathbf{3} \otimes \mathbf{3} \right)_\mathbf{1}  = \alpha_1 \beta_1 + \alpha_2 \beta_3 + \alpha_3 \beta_2 \, . 
\end{eqnarray}

Using this basis, the relevant fixed points for our model are
\begin{subequations}
\begin{eqnarray}
\tau_C &=& \omega \, , \qquad \qquad  Y_{\mathbf{3},\mathbf{3'}}^{(k)} \left(\tau_C\right)= \begin{pmatrix} \delta_0^{\text{mod}(k, 6)}  \\ \delta_2^{\text{mod}(k, 6)} \\ \delta_4^{\text{mod}(k, 6)}\end{pmatrix}  \, , \\
\tau_A &=& \frac{1}{2} + \frac{i}{2} \, , \qquad Y_\mathbf{3'}^{(4)}\left(\tau_A\right) = \begin{pmatrix} 0 \\ -1 \\ 1 \end{pmatrix} \, , \\
\tau_B &=& \frac{3}{2} + \frac{i}{2} \, ,\qquad  Y_\mathbf{3'}^{(2)}\left(\tau_B\right) = \begin{pmatrix} 1 \\ 1- \sqrt{6} \\ 1+ \sqrt{6} \end{pmatrix} \, , \\
\tau'_B &=& -\frac{1}{2} + \frac{i}{2} \, ,\qquad  Y_\mathbf{3'}^{(2)}\left(\tau'_B\right) = \begin{pmatrix} 1 \\ 1+ \sqrt{6} \\ 1- \sqrt{6} \end{pmatrix} \, .  
\end{eqnarray}
\end{subequations}
As for the modular forms we have, at the lowest weight in $S_4$,
\begin{equation}
Y_\mathbf{2}^{(2)}\left(\tau_C\right) = \begin{pmatrix} 0 \\ 1 \end{pmatrix} \, , \qquad Y_\mathbf{3}^{(2)}\left(\tau_C\right) = \begin{pmatrix} 0 \\ 1 \\0 \end{pmatrix}\, .
\end{equation}
Higher weight modular forms are obtained through the tensor products of lower-weight modular forms:
\begin{equation}\label{eq:HigherWeights}
Y^{(k)} = Y^{(k-2)} \otimes Y^{(2)}  = \bigotimes^{k/2} Y^{(2)} \, , 
\end{equation}
which can be easily computed for the cusp, and we see that, up to weight 12:
\begin{subequations}
\begin{eqnarray}
k=2 \, &:& \, \qquad Y^{(2)}_\mathbf{2}\left(\tau_C\right)  = \begin{pmatrix} 0 \\ 1 \end{pmatrix} \, , \qquad Y_\mathbf{3}^{(2)}\left(\tau_C\right) = \begin{pmatrix} 0 \\ 1 \\0 \end{pmatrix}\, , \\
k=4 \, &:& \, \qquad Y^{(4)}_\mathbf{2}\left(\tau_C\right)  = \begin{pmatrix} 1 \\ 0 \end{pmatrix} \, , \qquad Y_\mathbf{3}^{(4)}\left(\tau_C\right) = \begin{pmatrix} 0 \\ 0 \\1 \end{pmatrix}\, , \\
k=6 \, &:& \, \qquad Y^{(6)}_\mathbf{1}\left(\tau_C\right)  = \begin{pmatrix} 1  \end{pmatrix} \, , \qquad Y^{(6)}_\mathbf{1'}\left(\tau_C\right)  = \begin{pmatrix} 1  \end{pmatrix} \, , \qquad Y_\mathbf{3}^{(2)}\left(\tau_C\right) = \begin{pmatrix} 1 \\ 0 \\0 \end{pmatrix}\, , \\
k=8 \, &:& \, \qquad Y^{(8)}_\mathbf{2}\left(\tau_C\right)  = \begin{pmatrix} 0 \\ 1 \end{pmatrix} \, , \qquad Y_\mathbf{3}^{(8)}\left(\tau_C\right) = \begin{pmatrix} 0 \\ 1 \\0 \end{pmatrix}\, , \\
k=10 \, &:& \, \qquad Y^{(10)}_\mathbf{2}\left(\tau_C\right)  = \begin{pmatrix} 1 \\ 0 \end{pmatrix} \, , \qquad Y_\mathbf{3}^{(10)}\left(\tau_C\right) = \begin{pmatrix} 0 \\ 0 \\1 \end{pmatrix}\, , \\
k=12 \, &:& \, \qquad Y^{(12)}_\mathbf{1}\left(\tau_C\right)  = \begin{pmatrix} 1  \end{pmatrix} \, , \qquad Y^{(12)}_\mathbf{1'}\left(\tau_C\right)  = \begin{pmatrix} 1  \end{pmatrix} \, , \qquad Y_\mathbf{3}^{(2)}\left(\tau_C\right) = \begin{pmatrix} 1 \\ 0 \\0 \end{pmatrix}\, , 
\end{eqnarray}
\end{subequations}
where we only show the non-vanishing modular forms.  
It is clear that the pattern repeats, such that the non-vanishing modular forms at $k=2, 4, 6$ and $k=8, 10, 12$ are identical, respectively.  
Additionally, given Eq.~\eqref{eq:HigherWeights}, and that the modular forms of weight $6$ and $12$ are identical, then there is no difference in computing $Y^{(2)} \otimes Y^{(6)}$ and $Y^{(2)} \otimes Y^{(12)}$.  
Thus, at the cusp, it becomes obvious that the modular forms are given by:
\begin{subequations}
\begin{eqnarray}
Y_\mathbf{1}^{(k)} \left(\tau_C\right)&=& \begin{pmatrix} \delta_0^{\text{mod}(k, 6)} \end{pmatrix}  \, ,\qquad 
Y_\mathbf{1'}^{(k)} \left(\tau_C\right)= \begin{pmatrix} \delta_0^{\text{mod}(k, 6)} \end{pmatrix}  \, ,\\
Y_\mathbf{2}^{(k)}\left(\tau_C\right) &=& \begin{pmatrix} \delta_4^{\text{mod}(k, 6)}  \\ \delta_2^{\text{mod}(k, 6)}\end{pmatrix}  \, ,\qquad
Y_{\mathbf{3},\mathbf{3'}}^{(k)} \left(\tau_C\right)= \begin{pmatrix} \delta_0^{\text{mod}(k, 6)}  \\ \delta_2^{\text{mod}(k, 6)} \\ \delta_4^{\text{mod}(k, 6)}\end{pmatrix}  \, ,
\end{eqnarray}
\end{subequations}
assuming they exist in at a certain modular weight.  
This is useful for model building at fixed points, since it allows us to easily identify the shape of the modular forms of higher weights, without the need for actual computation.  

\section{Possible Corrections to the CSD($n$) Matrices}
\label{app:ExtraTerms}

At first glance, the assignments of the model shown in the main text appear more convoluted than necessary.  
Indeed, it is possible to find a seemingly simpler model, which includes all the terms and invariants we find in the main text.  
However, it is important to check if there are (non-negligible) corrections to any of the structures found for the relevant Yukawa matrices.  
In this Appendix, we show a simple example of this.  
Table~\ref{tab:messenger} shows one possible set of assignments for the fields that also lead to the structures found in the main text for $Y_u$ and $Y_d$.  
Namely, we find
\begin{subequations}
\begin{eqnarray}\label{eq:MassMats-2}
&Y_u =  
\begin{pmatrix} 
y_{uu} \epsilon_T^4 + y_{uu}' \epsilon_F^3 \epsilon_T & y_{uc} \epsilon_T^3 + y_{uc}' \epsilon_F^3 & y_{ut} \epsilon_T^2 \\
. & y_{cc} \epsilon_T^2 & y_{ct} \epsilon_T \\
.& . & y_{tt} 
\end{pmatrix}\, , & \\
&Y_d = 
\begin{pmatrix}
y_{dd} \epsilon_F^3 & y_{ds} \epsilon_F^2 \epsilon_T & y_{db} \epsilon_F \epsilon_T^2 \\
0 & y_{ss} \epsilon_F^2 & y_{sb} \epsilon_F \epsilon_T \\
0 & 0 & y_{bb} \epsilon_F
\end{pmatrix} \, \qquad 
Y_\ell = 
\begin{pmatrix}
y_{ee} \epsilon_F^3 & 0 & 0 \\
y_{\mu e} \epsilon_F^2 \epsilon_T & y_{\mu \mu} \epsilon_F^2 & 0 \\
y_{\tau e} \epsilon_F \epsilon_T^2 & y_{\tau \mu} \epsilon_F \epsilon_T & y_{\tau \tau} \epsilon_F
\end{pmatrix} \, .
&
\end{eqnarray}
\end{subequations}
as we do with the model in the main text.  
Moreover, the charges of the weightons ($\phi_F$, $\phi_T$) under $S_4^A$ will forbid higher-order corrections to these matrices.  

\begin{table}[h]
\centering
\renewcommand{\arraystretch}{1.2}
\begin{tabular}{|c| c | c c | c c | c c |}
\hline
Field 				& 	$SU(5)$	 				&	${S}_4^A$ 		&	$k_A$ 	&	${S}_4^B$ 		&	$k_B$ 		&	${S}_4^C$ 		& 	$k_C$ 		\\
\hline \hline
$F$ 					&	$\mathbf{5}$				& 	$\mathbf{1}$		&	$+\frac{1}{2}$	&	 $\mathbf{1}$		&	$  0$		&	$\mathbf{3}$		&	$  0$		\\
$T_1$				&	$\mathbf{10}$			& 	$\mathbf{1}$		&	$+1$\		&	 $\mathbf{1}$		&	$  0$		&	$\mathbf{1'}$		&	$  0$		\\
$T_2$				&	$\mathbf{10}$			& 	$\mathbf{1}$		&	$+\frac{1}{2}$	&	 $\mathbf{1}$		&	$  0$		&	$\mathbf{1'}$		&	$  0$		\\
$T_3$				&	$\mathbf{10}$			& 	$\mathbf{1}$		&	$  0$		&	 $\mathbf{1}$		&	$  0$		&	$\mathbf{1'}$		&	$  0$		\\
$N_A^c$				&	$\mathbf{1}$				& 	$\mathbf{1}$		&	$  4$		& 	$\mathbf{1}$		&	$  0$		&	$\mathbf{1}$		&	$  0$		\\
$N_B^c$				&	$\mathbf{1}$				& 	$\mathbf{1}$		&	$  0$		& 	$\mathbf{1}$		&	$+2$		&	$\mathbf{1}$		&	$  0$		\\
\hline
$\Phi_{AC}$			&	$\mathbf{1}$				&	$\mathbf{3}$		&	$  0$		&	$\mathbf{1}$		&	$  0$		&	$\mathbf{3}$		&	$  0$		\\
$\Phi_{BC}$			&	$\mathbf{1}$				&	$\mathbf{1}$		&	$  0$		&	$\mathbf{3}$		&	$  0$		&	$\mathbf{3}$		&	$  0$		\\
\hline
$\phi_{T}$			&	$\mathbf{1}$				&	$\mathbf{1}$		&	$-\frac{1}{2}$	&	$\mathbf{1}$		&	$  0$		&	$\mathbf{1}$		&	$  0$		\\
$\phi_{F}$			&	$\mathbf{1}$				&	$\mathbf{1}$		&	$-\frac{1}{2}$	&	$\mathbf{1}$		&	$  0$		&	$\mathbf{1}$		&	$+2$		\\
\hline
$H_5$				&	$\mathbf{5}$				&	$\mathbf{1}$		&	$ 0$			&	$\mathbf{1}$		&	$ 0$			&	$\mathbf{1}$		&	$ 0$			\\
$H_{\overline{5}}$		&  $\mathbf{\overline{5}}$		&	$\mathbf{1}$		&	$ 0$			&	$\mathbf{1}$		&	$ 0$			&	$\mathbf{1}$		&	$ 0$			\\
$H_{\overline{45}}$	&  $\mathbf{\overline{45}}$		&	$\mathbf{1}$		&	$ 0$			&	$\mathbf{1}$		&	$ 0$			&	$\mathbf{1}$		&	$ 0$			\\
\hline
\end{tabular}
\caption{\label{tab:messenger} A seemingly simpler assignment under the three modular symmetries, but which lead to non-negligible contributions which spoil the CSD($n$) structure.  
As in the main text, we do not show the messenger fields nor any necessary driving fields.}
\end{table}

As we can see from the weight assignments for the RH neutrinos, here the $SU(5)$ singlets will have bare mass terms, at the renormalizable level, contrary to what we see in Eq.~\eqref{eq:wM}.  
Mixed terms are still forbidden by the absence of the $Y_1^{(2)}$ modular form, needed to make a $N_A^c N_B^c$ term invariant.  

\begin{figure}[h!]
\centering
\begin{tikzpicture}
\begin{feynman}
\vertex (a1) {$F$};
\vertex[right=2cm of a1, dot] (a2) {};
\vertex[right=4cm of a2, dot] (a3) {};
\vertex[right=4cm of a3, dot] (a4) {};
\vertex[right=2cm of a4] (a5) {$N_A^c$};
\vertex[above=2.25cm of a2] (b1) {\( H_5 \)};
\vertex[above=2.25cm of a3] (b2) {\( \phi_T \)};
\vertex[above=2.25cm of a4] (b3) {\(\Phi_{AC} \)};
\vertex[above right =0.6cm of a2] (c1) {$Y_\mathbf{1}^{(0)}$};
\vertex[above right =0.6cm of a3] (c1) {$Y_\mathbf{1}^{(0)}$};
\vertex[above right =0.6cm of a4] (c1) {$Y_\mathbf{3'}^{(4)}$};
\diagram* {
{[edges=fermion]
(a1) -- (a2) [dot],
(a4) -- (a5),
},
(a2) -- [scalar] (b1),
(a3) -- [scalar] (b2),
(a4) -- [scalar] (b3) ,
(a2) -- [insertion={[size=3pt]0.53}, edge label'= $  \begin{pmatrix} \mathbf{1}_{-\frac{1}{2}} \text{,} &   \mathbf{3}_0  \end{pmatrix} \quad  \begin{pmatrix} \mathbf{1}_{\frac{1}{2}} \text{,} &   \mathbf{3}_0  \end{pmatrix} $ ] (a3) [dot],
(a3) -- [insertion={[size=3pt]0.5}, edge label'= $  \begin{pmatrix} \mathbf{1}_{0} \text{,} &   \mathbf{3}_0  \end{pmatrix} \quad  \begin{pmatrix} \mathbf{1}_{0} \text{,} &   \mathbf{3}_0  \end{pmatrix} $ ] (a4) [dot],
(a3) -- [scalar] (a4) [dot],
(a4) -- [scalar] (a5)
};
\end{feynman}
\end{tikzpicture}
\begin{tikzpicture}
\begin{feynman}
\vertex (a1) {$F$};
\vertex[right=2cm of a1, dot] (a2) {};
\vertex[right=4cm of a2, dot, red] (a3) {};
\vertex[right=4cm of a3, dot] (a4) {};
\vertex[right=2cm of a4] (a5) {$N_A^c$};
\vertex[above=2.25cm of a2] (b1) {\( H_5 \)};
\vertex[above=2.25cm of a3] (b2) {\( \phi_F \)};
\vertex[above=2.25cm of a4] (b3) {\(\Phi_{AC} \)};
\vertex[above right =0.6cm of a2] (c1) {$Y_\mathbf{1}^{(0)}$};
\vertex[above right =0.6cm of a3] (c1) {$Y_\mathbf{3'}^{(2)}$};
\vertex[above right =0.6cm of a4] (c1) {$Y_\mathbf{3'}^{(4)}$};
\diagram* {
{[edges=fermion]
(a1) -- (a2) ,
(a4) -- (a5),
},
(a2) -- [scalar] (b1),
(a3) -- [scalar] (b2),
(a4) -- [scalar] (b3) ,
(a2) -- [insertion={[size=3pt]0.53}, edge label'= $  \begin{pmatrix} \mathbf{1}_{-\frac{1}{2}} \text{,} &   \mathbf{3}_0  \end{pmatrix} \quad  \begin{pmatrix} \mathbf{1}_{\frac{1}{2}} \text{,} &   \mathbf{3}_0  \end{pmatrix} $ ] (a3) ,
(a3) -- [insertion={[size=3pt]0.5}, edge label'= $  \begin{pmatrix} \mathbf{1}_{0} \text{,} &   \mathbf{3}_0  \end{pmatrix} \quad  \begin{pmatrix} \mathbf{1}_{0} \text{,} &   \mathbf{3}_0  \end{pmatrix} $ ] (a4) ,
(a3) -- [scalar] (a4) ,
(a4) -- [scalar] (a5)
};
\end{feynman}
\end{tikzpicture}
\caption{\label{fig:diags} The diagrams leading to the desired (top) and undesired (bottom) terms for $w_D$.  The modular forms refer to $S_4^A$, as the remaining are trivial.  The messengers are fermionic $SU(5)$ singlets, represented by the $S_4^A$ and $S_4^C$ representations and weights, and assumed to transform trivially under $S_4^B$.}
\end{figure}
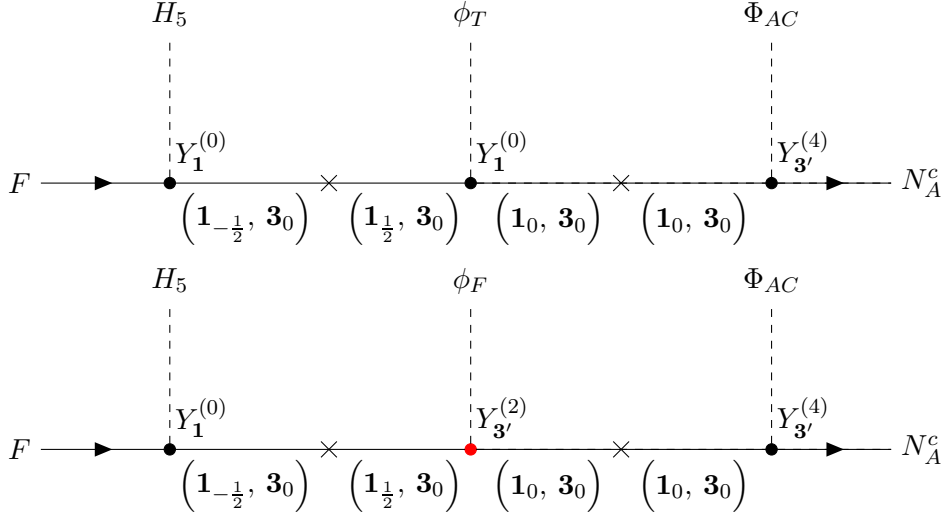

The last ingredient needed is to reproduce the Dirac mass matrix compatible with the CSD($n$) structure.  
There, we find that the invariants 
\begin{equation}\label{eq:wD2}
w_D \supset H_5 \left\{ a Y_{3'}^{(4)}(\tau_A) F \left(\dfrac{\phi_T}{\Lambda} \dfrac{\left< \Phi_{AC}\right>}{\Lambda} \right) N_A^c + b Y_{3'}^{(2)}(\tau_B) F \left(\dfrac{\phi_T}{\Lambda} \dfrac{\left< \Phi_{BC}\right>}{\Lambda} \right) N_B^c \right\}
\end{equation}
are present, as required, similarly to Eq.~\eqref{eq:wD}. 
On the other hand, we can now make another set of invariants, by replacing $\phi_T$ with $\phi_F$.  
If we follow the same tensor product contractions as the terms in $w_D$, these would be forbidden by the absence of the $Y_1^{(2)}$ modular form.  
However, we can contract the term as (we show the example for $F N_A^c$)
\begin{equation}\label{eq:nuissance}
\Bigg(\bigg(\left( Y_\mathbf{3'}^{(2)}(\tau_C) \otimes (F\phi_F) \right)_\mathbf{3} \otimes \left<\Phi_{AC}\right> \otimes Y_\mathbf{3}^{(4)}(\tau_A) \bigg)_\mathbf{1'} \otimes \left(N_A^c\right)_\mathbf{1'} \Bigg)_1 \, , 
\end{equation}
which spoil the $Y_D$ structure.  
At first glance, we could argue that we could choose the model's messengers such that the terms in Eq.~\eqref{eq:wD2} are present, but those of Eq.~\eqref{eq:nuissance} are absent.  
However, we checked that, for the simplest choice of messengers, both terms are necessarily present.  
This is shown in the diagrams of Fig.~\ref{fig:diags}, where we can clearly see that the same set of messengers lead to the existence of both terms.  
The assignments of the model presented in the main text are mostly motivated to eliminate these terms, and keep an unspoiled $Y_D$, such that the corrections to the CSD($n$) structure arise solely from the non-diagonal $Y_\ell$, in a suppressed manner.

\bibliographystyle{JHEP}
\bibliography{LMSU5.bib}

\end{document}